\documentclass[aps,pra,twocolumn,longbibliography]{revtex4-1}

\usepackage[latin1]{inputenc}
\usepackage{amsmath}
\usepackage{amssymb}
\usepackage{amsfonts}
\usepackage{wasysym}
\usepackage{bbold}
\usepackage{graphicx}
\usepackage{times}
\usepackage{color}

\newcommand{\ket}[1]{\vert{#1}\rangle}
\newcommand{\bra}[1]{\langle{#1}\vert}
\newcommand{\norm}[1]{\|{#1}\|}

\newcommand{\av}[1]{\mathbb{E}({#1})}
\newcommand{\var}[1]{\mathrm{Var}({#1})}
\newcommand{\gen}{\mathcal{L}}

\newcommand{\rme}{\mathrm{e}}
\newcommand{\rmi}{\mathrm{i}}
\newcommand{\rmd}{\mathrm{d}}
\newcommand{\tr}{\mathrm{tr}}
\newcommand{\id}{\mathbb{1}}
\newcommand{\jd}{\mathbb{J}}

\begin{document}

\title{Decoherence enhances performance of quantum walks applied to graph isomorphism testing}

\author{M.~Bruderer}
\author{M.~B.~Plenio}
\address{Institut für Theoretische Physik, Albert-Einstein Allee 11, Universität Ulm, 89069 Ulm, Germany}

\date{\today}

\begin{abstract}
Computational advantages gained by quantum algorithms rely largely on the coherence of quantum devices and
are generally compromised by decoherence. As an exception, we present a quantum algorithm for graph
isomorphism testing whose performance is optimal when operating in the partially coherent regime, as opposed
to the extremes of fully coherent or classical regimes. The algorithm builds on continuous-time quantum
stochastic walks (QSWs) on graphs and the algorithmic performance is quantified by the distinguishing power (DIP)
between non-isomorphic graphs. The QSW explores the entire graph and acquires information about the underlying
structure, which is extracted by monitoring stochastic jumps across an auxiliary edge. The resulting
counting statistics of stochastic jumps is used to identify the spectrum of the dynamical generator of the QSW,
serving as a novel graph invariant, based on which non-isomorphic graphs are distinguished. We provide specific
examples of non-isomorphic graphs that are only distinguishable by QSWs in the presence of decoherence.
\end{abstract}

\maketitle


\section{Introduction}

The main objective of quantum information processing is to enhance the computational performance of
quantum devices over comparable classical devices by exploiting coherent quantum effects. Incoherent
or even dissipative processes therefore generally pose major obstacles to physical implementations of
quantum computing~\cite{bennett2000quantum}. That being said, it is known that incoherent effects can
be utilized to accomplish specific tasks such as entanglement
generation~\cite{plenio1999cavity,plenio2002entangled,contreras2008entanglement} and quantum
teleportation~\cite{bose1999proposal}. Another example for beneficial effects of decoherence is dephasing-enhanced
transport found in systems as diverse as light-harvesting complexes~\cite{plenio2008dephasing,mohseni2008environment},
structured waveguide arrays~\cite{caruso2015fast}, one-dimensional conductors~\cite{mendoza2013dephasing}
and arrays of quantum dots~\cite{Contreras-NJP-2014}. At the level of quantum algorithms for specific
mathematical problems, however, it seems to be an established paradigm that algorithms relying on purely
coherent quantum dynamics always outperform partially decoherent approaches~\cite{mosca2009quantum,childs2010quantum}.
As a rare counterexample, algorithmic applications of decoherence have been reported in the
context of quantum walks~\cite{kendon2003decoherence,kendon2007decoherence}.

In this paper we present a quantum algorithm whose performance is optimal when operating in the partially
coherent regime and thus benefits from decoherence. The algorithm solves instances of a clearly defined
mathematical task, namely the graph isomorphism (GI) problem~\cite{Kobler-1993}. The GI problem is central
to graph theory and consists of testing if two graphs are isomorphic,~i.e.,~if one graph can be mapped to
the other by a relabeling of vertices. As of today, no efficient (polynomial-time) algorithm for the GI problem
is known in full generality. However, for the vast majority of graphs, the problem can be solved efficiently in
practice~\cite{mckay2014practical}, and it has been recently shown that the GI problem can be solved in
quasipolynomial-time in the worst case~\cite{babai2016graph}.

Here we consider a physically motivated approach towards solving the GI problem for certain classes of graphs.
As often for GI testing, the performance of the algorithm is quantified by the distinguishing power (DIP) between
non-isomorphic graphs, in contrast to runtime efficiency. The proposed algorithm for GI testing is formulated in
terms of continuous-time quantum stochastic walks (QSWs) on graphs, whose dynamics spans the entire range between
fully coherent quantum walks and classical random walks~\cite{Whitfield-PRA-2010}. QSWs allow us to study the
workings of our algorithm under decoherence and to explore its quantum-to-classical transition. Similar to the results
in Refs.~\cite{kendon2003decoherence,kendon2007decoherence}, we find that already a small amount of decoherence enhances
the performance of our quantum algorithm.

\begin{figure}[h]
\centering
\includegraphics[width=0.95\columnwidth]{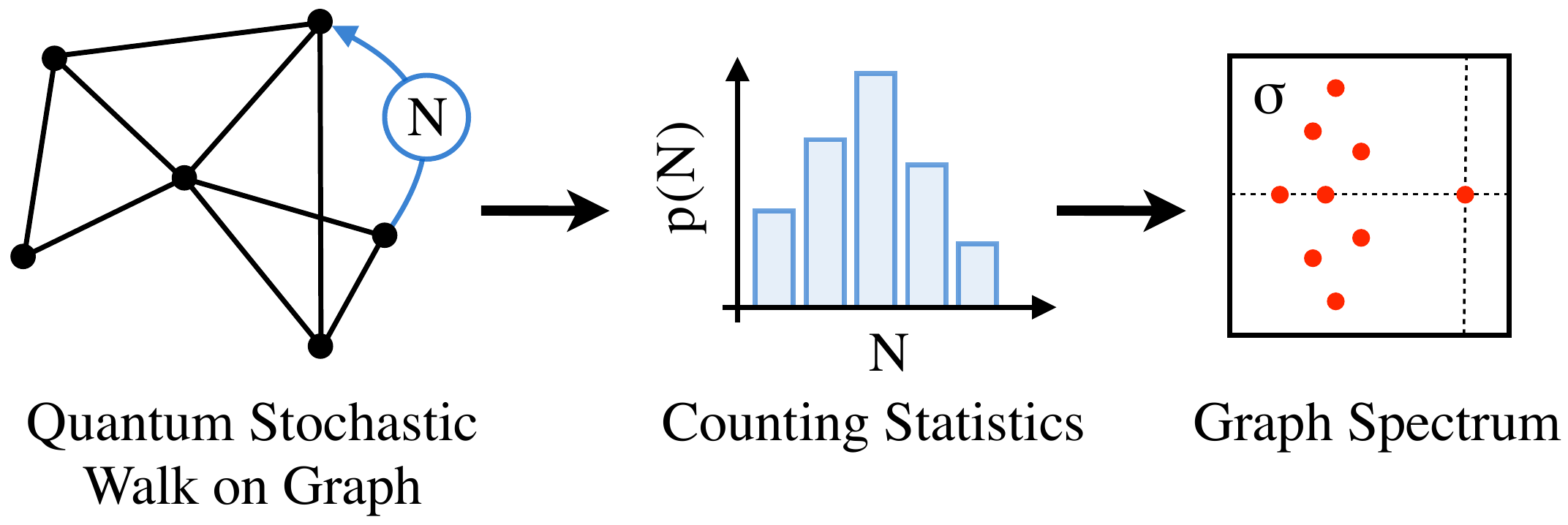}
\caption{Proposed algorithm for graph isomorphism testing: A quantum stochastic walk (QSW) on the
graph relaxes to the steady state. The number of jumps $N$ per time interval across an auxiliary
edge is monitored, yielding the counting statistics $p(N)$. The distribution $p(N)$ is used to
determine the complex spectrum $\sigma$ of the generator of the QSW. Graphs are compared by
their spectra, where different spectra indicate non-isomorphism.}
\label{scheme}
\end{figure}

The GI problem enjoys renewed popularity and several quantum (or quantum-inspired) algorithms for GI
testing have been presented recently. As shown in Ref.~\cite{rudolph2002constructing}, absorption spectra
of exciton Hamiltonians with graph-structured interactions can be used to distinguish graphs. Algorithms
for GI testing based on multi-particle quantum
walks, relying on comparisons between evolution operators or occupation probabilities, have been proposed in
Refs.~\cite{shiau2005physically,douglas2008classical,gamble2010two,berry2011two,rudinger2012noninteracting},
where the DIP depends on modifications of local phases and whether the evolution is discrete or
continuous in time~\cite{rudinger2013comparing,mahasinghe2015phase}. Ideas originating from quantum
walks have also been adapted to design classical algorithms for the GI problem~\cite{emms2009coined,tamascelli2014quantum}.
Apart from their theoretical appeal, both discrete- and continuous-time quantum walks have been
experimentally implemented on various physical platforms, including NMR systems~\cite{ryan2005experimental},
trapped ions~\cite{schmitz2009quantum,zahringer2010realization} and photonic
implementations~\cite{perets2008realization,schreiber2010photons,aspuru2012photonic}.

Our algorithm is conceptually different from existing approaches in that both the initialization
and readout are continuous processes. GI testing is achieved with the desired precision by continuously monitoring
local fluctuations of QSWs on the level of random trajectories, described by the counting statistics of stochastic
jumps~\cite{garrahan2010thermodynamics,schaller2014open}. Another feature is that, in the tradition of spectral
graph theory~\cite{cvetkovic1980,brouwer2011spectra}, the complex spectrum of the dynamical generator of the QSW
serves as a graph invariant. In addition to its use as part of our quantum algorithm, this novel graph invariant
may also serve as the basis for quantum-inspired algorithms.

Figure~\ref{scheme} illustrates the essential steps of the algorithm: For the initialization, an auxiliary edge is
connected to two arbitrary vertices of the graph and the QSW is allowed to relax to the steady state. Subsequently,
a counting device monitors stochastic jumps of the QSW across the auxiliary edge. During this continuous readout process,
information about the graph structure is accumulated and encoded in the counting statistics $p(N)$ of the number of jumps $N$
per time interval. Finally, the information contained in the distribution $p(N)$ is used to determine the spectrum $\sigma$
of the generator of the QSW (as defined later). Isomorphism of two graphs $G$ and $G^\prime$ is tested by comparison of
their spectra $\sigma$ and $\sigma^\prime$, where different spectra imply non-isomorphic graphs.

The individual steps of the algorithm are explained in the rest of the paper. After preliminaries about graphs, we initially
focus on the generator of the QSW as a matrix representation for graphs and analyze the spectrum of the generator for different
levels of decoherence.


\section{Graph invariants from quantum stochastic walks}

We consider undirected graphs $G=(V,E)$ consisting of the vertex set $V$ and edge set $E$,
where $(i,j)\in E$ denotes an edge between vertices $i, j\in V$ and $n=\vert V\vert$ is the
number of vertices. The graphs are connected with neither multiple edges nor self-loops.
The graph structure is encoded in the adjacency matrix $A$, with entries
$A_{ij} = 1$ if $(i,j)\in E$ and $A_{ij} = 0$ otherwise, or described by the Laplacian matrix
$L = D - A$. The degree matrix $D$ with entries $D_{ii}=d_i$ is diagonal, where the degree
$d_i$ is the number of edges connected to vertex~$i$.
Other common matrix representations are the signless Laplacian $\vert L\vert = D + A$
and the adjacency matrix of the complement $\overline{A} = \jd - A - \id$, with $\id$ the identity
and  $\jd$ the all-ones matrix. The GI problem in terms of matrix representations is equivalent to
deciding if two adjacency matrices $A$ and $A^\prime$ represent isomorphic graphs,~i.e.,~whether or
not there exists a permutation matrix $\Pi$ such that $A = \Pi^{-1}A^\prime\Pi$ holds.

An important concept of GI testing are graph invariants, defined as quantities $I(G)$
such that $I(G) = I(G^\prime)$ if $G$ and $G^\prime$ are isomorphic~\cite{corneil1980theoretical,mckay2014practical}.
Graph invariants can be constructed through matrix representations $M$ of graphs, where $M$
stands for $A, L$ or other representations. Important invariants are the characteristic polynomial
$P^M(\mu) = \det[\mu\id - M]$ and the spectrum $\sigma^M$ consisting of the roots of $P^M(\mu)$. We
denote $\sigma^M$ as $M$-spectrum and call two graphs $M$-cospectral if their $M$-spectra coincide.
The coefficients of $P^M(\mu)$ are directly related to cycles and spanning trees of graphs~\cite{cvetkovic1980},
and the majority of graphs are identifiable by a single spectrum or combinations of different
spectra~\cite{van2003graphs,haemers2004enumeration,wilson2008study}.

We now turn to the generator of QSWs, which is an alternative matrix representation of the graph $G$, and
discuss the corresponding graph invariants.
QSWs on graphs account for both stochastic jumps and quantum tunneling between connected vertices~\cite{Whitfield-PRA-2010}.
For ensemble averages over trajectories, QSWs are described by an $n\times n$ density matrix $\rho$ whose time
evolution is governed by the quantum master equation 
\begin{equation}\label{lindblad}
	\frac{\rmd \rho(t)}{\rmd t} = \gen^\omega\rho(t) = \big[\omega\gen^{\rm qm} + (1-\omega)\gen^{\rm cl}\big]\rho(t)\,.
\end{equation}
The generator $\gen^\omega$ is decomposed into $\gen^{\rm qm}$ and $\gen^{\rm cl}$ such that the coherence
$\omega\in[0,1]$ parametrizes the quantum-to-classical transition. In terms of quantum walks under decoherence the parameter
$1-\omega$ quantifies the strength of adverse environmental effects. The generator $\gen^{\rm qm}$ of the coherent
dynamics acts on the density matrix as $\gen^{\rm qm}\rho =  -\rmi[A,\rho]$
and the generator $\gen^{\rm cl}$ of the classical dynamics is the Lindblad dissipator
\begin{equation}\label{stgen}
\gen^{\rm cl}\rho = \sum_{i,j}A_{ij}\Big(\Upsilon_{ij}\rho\Upsilon_{ij}^\dagger
	- \frac{1}{2}\{\Upsilon_{ij}^\dagger \Upsilon_{ij},\rho\}\Big),
\end{equation}
where the operators $\Upsilon_{ij} = \ket{j}\!\bra{i}$ map state $\ket{i}$ to $\ket{j}$. The dissipator
$\gen^{\rm cl}$ induces stochastic jumps between vertices and is closely related to the Laplacian $L$,
the generator of classical random walks~\cite{mirzaev2013laplacian}. Dephasing operators
$\Upsilon_{ii} = \ket{i}\!\bra{i}$ are not contained in $\gen^{\rm cl}$ since $A_{ii}=0$; however,
dephasing terms occur for graphs with self-loops.

The generator $\gen^\omega$ is a matrix representation in the same sense as the adjacency matrix $A$.
As a linear superoperator in Liouville space, $\gen^\omega$ is represented by an $n^2\times n^2$
matrix in the dyadic basis $\{\vert{i}\rangle\!\langle{j}\vert\}$ (with $i,j = 1,\ldots,n$) acting on the vector
$\rho$ with $n^2$ components. In this representation we obtain explicit expressions for the characteristic
polynomial $P^\omega(\nu)=\det[\nu\id - \gen^\omega]$ and the spectrum $\sigma^\omega=\{\nu_1,\ldots,\nu_{n^2}\}$.
The eigenvalues $\nu_i$ are either real-valued or come in complex conjugate pairs such that
$P^\omega(\nu) = \nu^{n^2} + q_{n^2-1}\nu^{n^2-1} + \cdots + q_1\nu + q_0$ has real coefficients $q_i$.
For a detailed understanding of the spectrum of $\gen^\omega$ ($\omega$-spectrum for short) as a graph
invariant we consider the extreme limits $\omega = \{0,1\}$, thereby relating the $\omega$-spectrum to the
spectrum of the adjacency and Laplacian matrix.

The generator $\gen^{\rm cl}$ of the classical random walk ($\omega = 0$) is applied to basis
vectors $\{\vert{i}\rangle\!\langle{j}\vert\}$ to determine its structure. Considering separately
occupations $\ket{i}\!\bra{i}$ and coherences $\ket{i}\!\bra{j}$ with $i\neq j$
we obtain
\begin{equation}\label{blockform}
\begin{split}
	\gen^{\rm cl}\ket{i}\!\bra{i} &= \sum_\ell A_{i\ell}\ket{\ell}\!\bra{\ell} - d_i\ket{i}\!\bra{i}\,,\\
	\gen^{\rm cl}\ket{i}\!\bra{j} &= -\frac{1}{2}(d_i + d_j)\ket{i}\!\bra{j}\,.
\end{split}	
\end{equation}
Equations~(\ref{blockform}) show that $\gen^{\rm cl}$ does not mix occupations and coherences, and is
therefore block diagonal in the basis $\{\vert{i}\rangle\!\langle{j}\vert\}$. The block acting on
occupations, in matrix form, is identical to the negative of the Laplacian $L$, whereas the coherences
are eigenvectors of $\gen^{\rm cl}$ with eigenvalues $-\frac{1}{2}(d_i + d_j)$. The spectrum
$\sigma^{\rm cl} = \{-\lambda_1,\ldots,-\lambda_{n},-\frac{1}{2}(d_i + d_j)\}$ is comprised of the spectra
of both blocks, where $\lambda_i$ are eigenvalues of the Laplacian $L$ and $i,j = 1,\ldots,n$
with $i\neq j$.
The generator $\gen^{\rm qm}$ of the quantum walk ($\omega = 1$) has a purely imaginary spectrum.
Since $\gen^{\rm qm}$ is essentially the commutator between $A$ and $\rho$ the eigenvectors of $\gen^{\rm qm}$
are of the form $\ket{\alpha_i}\!\bra{\alpha_j}$. Here, $\ket{\alpha_i}$ with $i=1,\ldots,n$ are the
eigenvectors with real eigenvalues $\alpha_i$ of the (real symmetric) adjacency matrix $A$.
We obtain $\gen^{\rm qm}\ket{\alpha_i}\!\bra{\alpha_j} = -\rmi(\alpha_i-\alpha_j)\ket{\alpha_i}\!\bra{\alpha_j}$
and the corresponding spectrum $\sigma^{\rm qm} =\{\rmi(\alpha_i-\alpha_j)\}$ with $i,j = 1,\ldots,n$.

Generally, the $\omega$-spectrum is a non-linear interpolation between $\sigma^{\rm cl}$ and $\sigma^{\rm qm}$,
parametrized by the coherence $\omega$. While analytic expressions for $\sigma^\omega$ may be obtained for simple
cases we show that the eigenvalues $\nu_i$ are finite and hence well defined for all values of $\omega$. First note
that the spectral radius $\varrho^\omega = \max_i\vert\nu_i\vert$ is bounded as $\varrho^\omega\leq\norm{\gen^\omega}$ 
for any matrix norm $\norm{\,\cdot\,}$. From the properties of norms follows that
$\norm{\gen^\omega}\leq \omega\norm{\gen^{\rm qm}} + (1-\omega)\norm{\gen^{\rm cl}}$,
implying that the spectral radius $\varrho^\omega$ and hence all eigenvalues $\nu_i$ are indeed bounded by
the finite expression $\norm{\gen^{\rm qm}} + \norm{\gen^{\rm cl}}$. For an explicit upper bound one can use
the Frobenius norm, defined by $\norm{M}^2=\sum_i\vert\mu_i\vert^2$ for normal matrices $M$ with eigenvalues
$\mu_i$. A more detailed characterization of $\sigma^\omega$ is provided in Ref.~\cite{wielandt1955eigenvalues}.


\section{Higher distinguishing power for partial coherence}

It is clear from the previous analysis that the $\omega$-spectrum for coherences $\omega=\{0,1\}$ is fully determined
by the $A$- and $L$-spectra, and the degrees $d_i$. Thus, the $\omega$-spectrum has the same DIP as these invariants
in the classical and fully coherent regimes; this is however not the case for intermediate coherence. The fundamental
yet simple reason is that the spectrum of $\omega\gen^{\rm qm} + (1-\omega)\gen^{\rm cl}$ is not identical to
$\omega\sigma^{\rm qm} + (1-\omega)\sigma^{\rm cl}$, except when $\gen^{\rm qm}$ and $\gen^{\rm cl}$ commute.
Consequently, the $\omega$-spectra of two non-isomorphic graphs may be different even if they have identical
traditional graph invariants,~i.e.,~the degrees $d_i$ and the spectra of $A, L, \overline{A}$ and $\vert L\vert$.
Conversely, two graphs for which the degrees $d_i$ and the $A$- and $L$-spectra are different always have different
$\omega$-spectra.

\begin{figure}[t]
\flushright
\includegraphics[width=245pt]{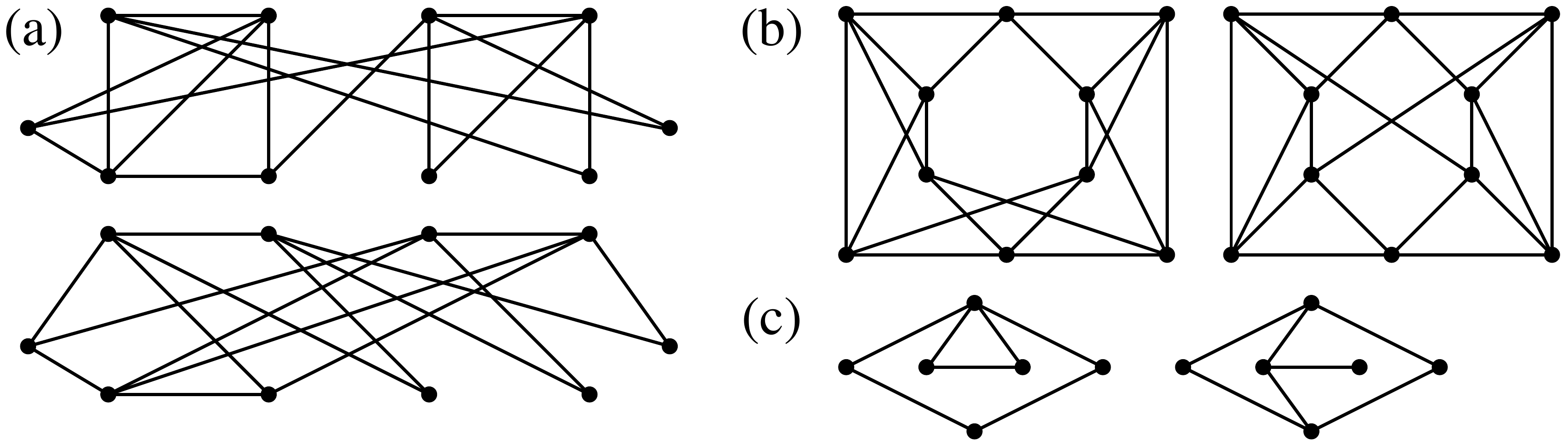}\vspace{15pt}\\
\includegraphics[width=120pt]{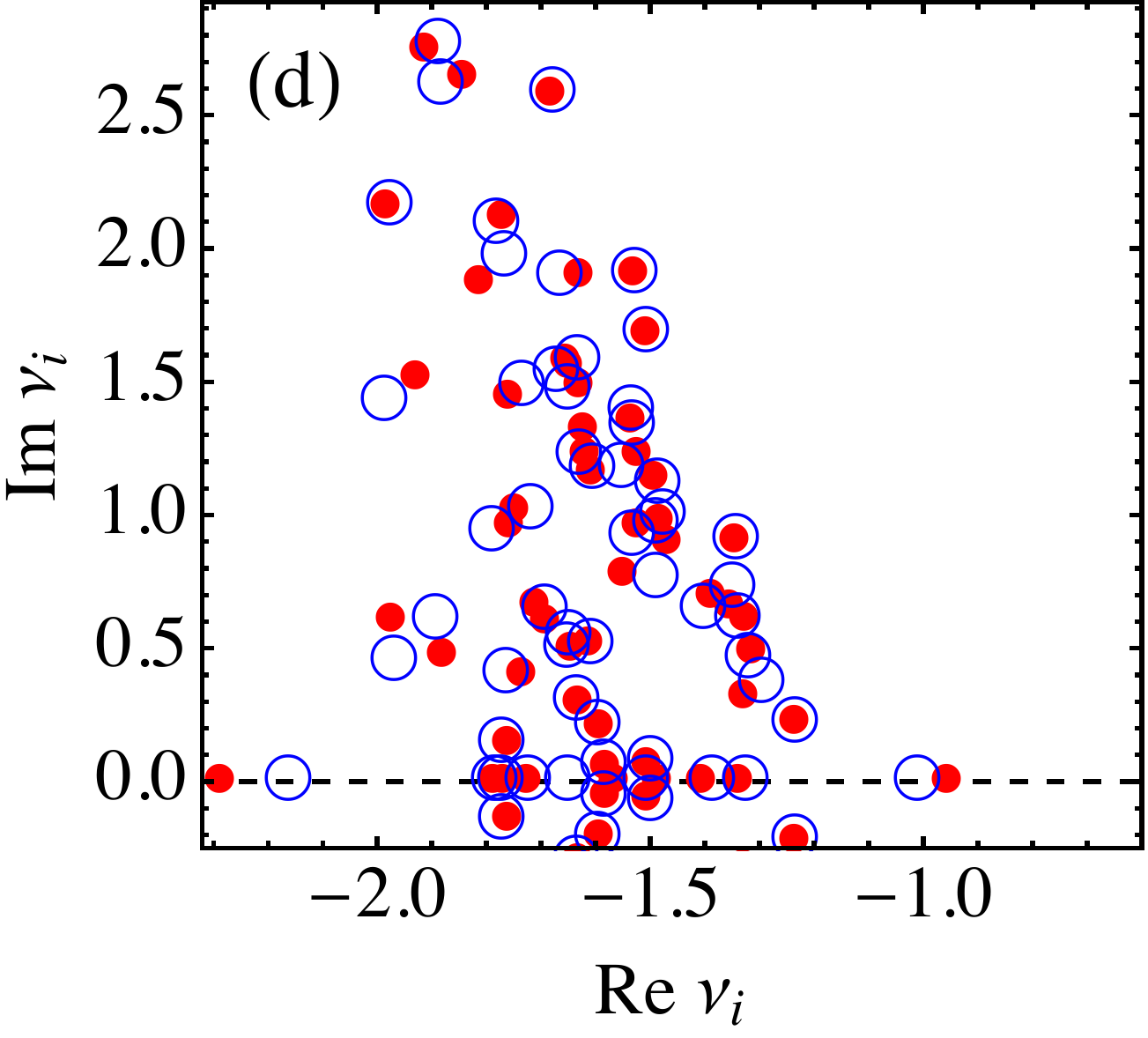}\hspace{3pt}
\includegraphics[width=120pt]{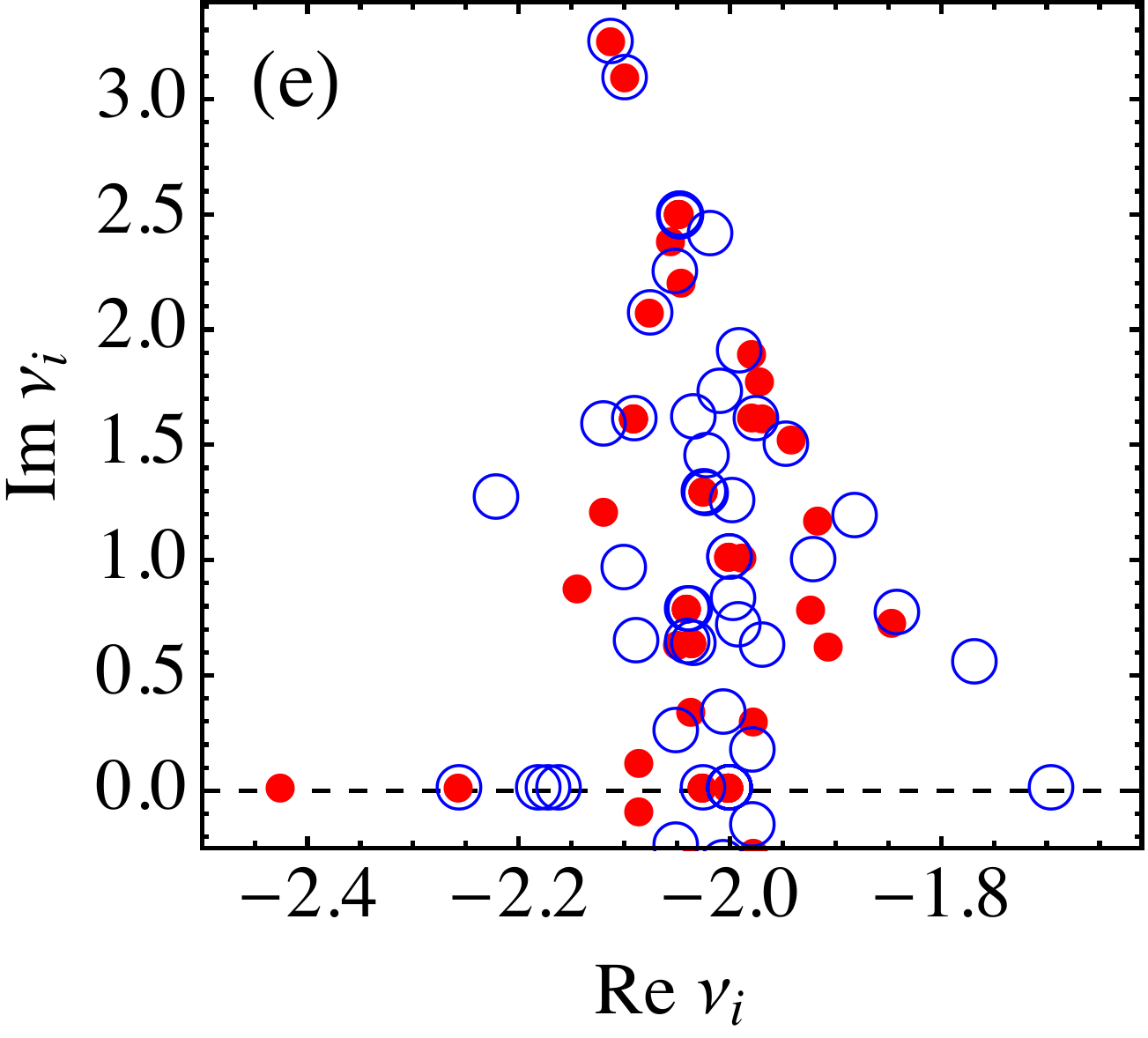}
\caption{Pairs of non-isomorphic graphs demonstrate the higher DIP of the $\omega$-spectrum. The pairs
in (a) and (b) are cospectral with respect to traditional matrix representations, but the corresponding
$\omega$-spectra in (d) and (e) are clearly distinct for intermediate coherence $\omega = 1/2$. The
eigenvalues of the individual graphs (marked by $\bullet$ and $\Circle$) are distributed symmetrically
about the real axis (dashed line). The pair of graphs in (c) is $L$-cospectral, but distinguishable
by the $\omega$-spectrum even in the classical regime $\omega=0$.}
\label{graph}
\end{figure}

The higher DIP for partially coherent QSWs is demonstrated by non-isomorphic graphs with different
$\omega$-spectra, but indistinguishable by traditional invariants~\cite{van2003graphs,haemers2004enumeration}.
The first example is provided by the pair of graphs in Fig.~\ref{graph}(a), cospectral with respect
to traditional matrix representations. Figure~\ref{graph}(d) shows the clearly distinct $\omega$-spectra
of the graphs for intermediate coherence $\omega=1/2$. Therefore, spectra of QSWs under the influence of
decoherence allow us to distinguish more graphs than QSWs in the classical and fully coherent regimes. The
next non-isomorphic pair in Fig.~\ref{graph}(b) consists of $A$-cospectral \emph{regular} graphs, having
identical degrees $d_i\equiv d$ for all vertices, for which the representations $A, L, \overline{A}$ and $\vert L\vert$
are equivalent with regard to graph spectra~\cite{van2003graphs}. Despite their high symmetry, the graphs have
also clearly distinct $\omega$-spectra, shown in Fig.~\ref{graph}(e).

As for standard spectral methods, the DIP of $\omega$-spectra is limited for graphs with highly degenerate eigenvalues.
Known examples are \emph{strongly regular} graphs~\cite{brouwer2011spectra}, whose adjacency matrix $A$ has only three
distinct eigenvalues for any number of vertices $n$. An example of a $\omega$-cospectral pair of non-isomorphic
strongly regular graphs are the Shrikhande graph and the lattice graph $L_2(4)$, having $16$ vertices and identical
$\omega$-spectra with only $12$ distinct eigenvalues. Several algorithms based on quantum walks are capable of distinguishing
specific families of strongly regular graphs (see Ref.~\cite{mahasinghe2015phase} for a comparison).

The DIP further reveals that classical random walks are not exactly equivalent to the classical limit
of QSWs, generated by the Laplacian $L$ and the dissipator $\gen^{\rm cl}$, respectively. The higher
DIP of the $\omega$-spectrum, compared to the $L$-spectrum, even persists in the classical regime
$\omega = 0$ because of its dependence on the degrees $d_i$. This is exemplified by the pair of graphs
in Fig.~\ref{graph}(c) with different degrees $d_i$, which are $L$-cospectral but distinguishable by the
$\omega$-spectrum for $\omega=0$.

\begin{figure}[t]
\centering
\includegraphics[width=130pt]{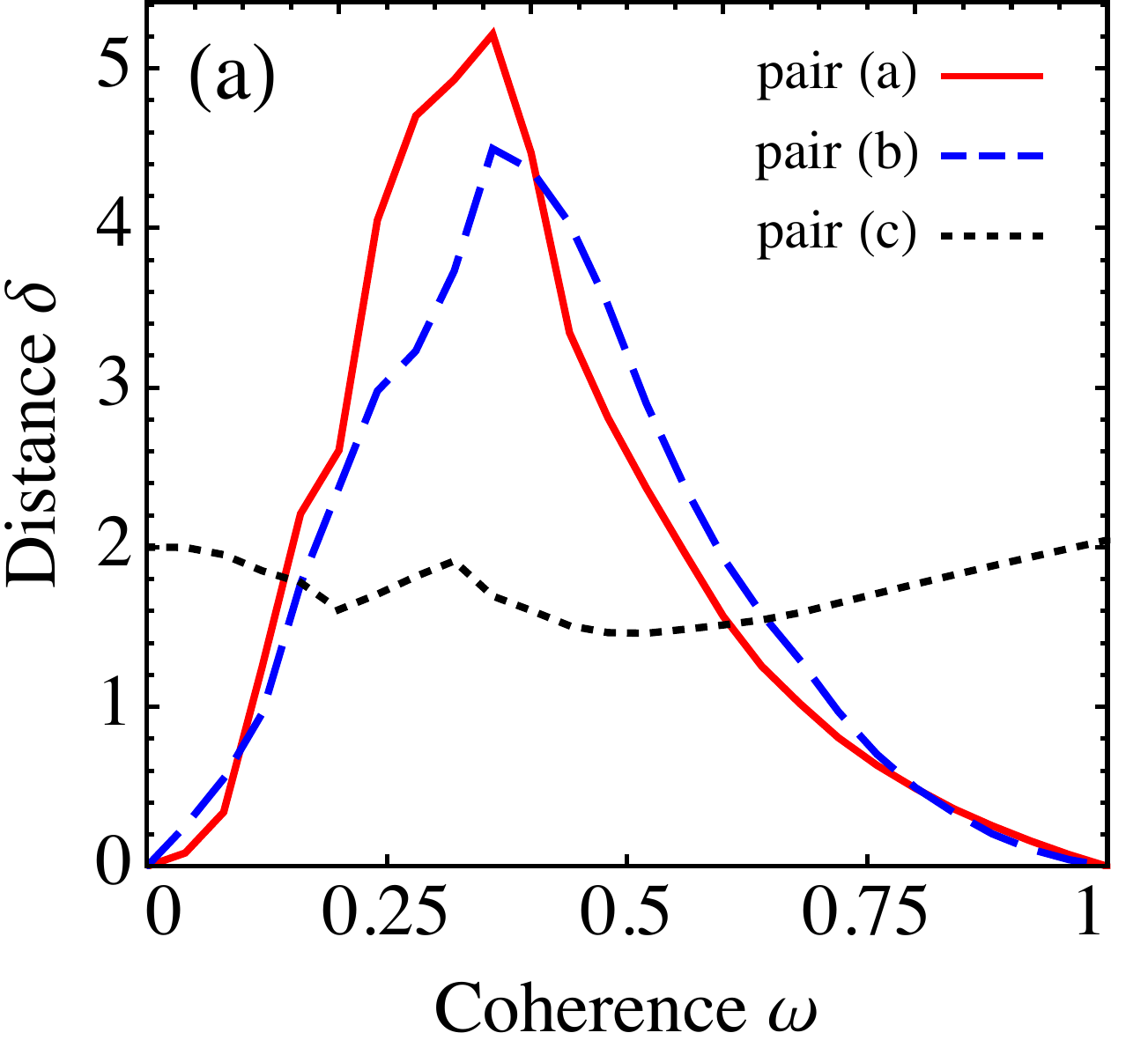}
\raisebox{20pt}{\includegraphics[width=112pt]{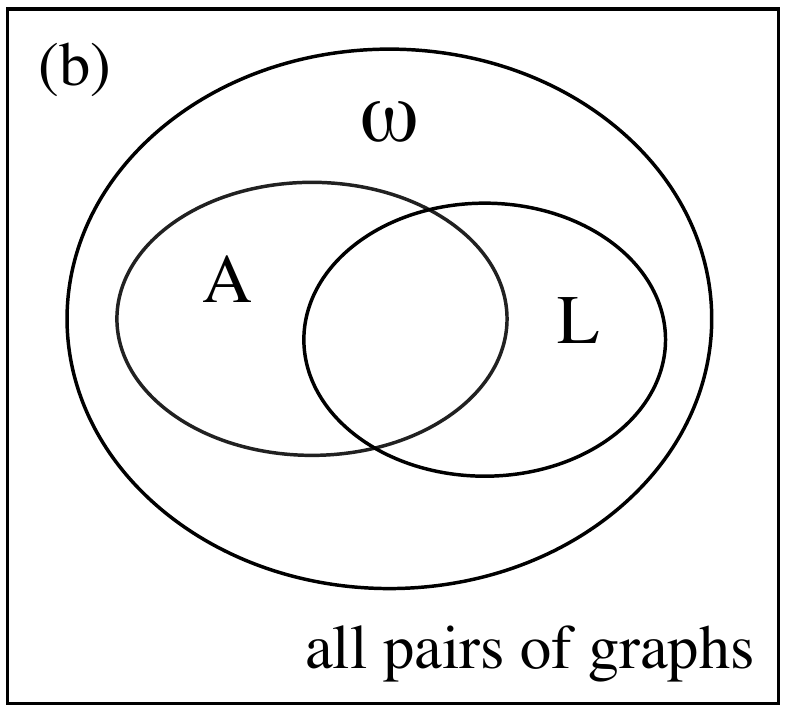}}
\caption{(a)~The distance $\delta$ between $\omega$-spectra depending on the coherence $\omega$
for the pairs of graphs in Fig.~\ref{graph}. For pairs~(a) and (b), the distance $\delta$ is peaked
for intermediate coherence $\omega$, whereas $\delta=0$ in the classical and fully coherent limits.
For the $L$-cospectral pair~(c), the distance $\delta$ (rescaled) is always nonzero, even in the
classical limit $\omega=0$.
(b)~Pairs of graphs categorized into classes according to distinguishability with respect to different
spectra. The class of pairs distinguishable by $\omega$-spectra includes the classes defined by $A$-
and $L$-spectra, and additional pairs not contained in these classes.}
\label{diagram}
\end{figure}

An obvious question is how the DIP depends on the level decoherence. We quantify the difference between the
spectra of two graphs by the distance measure $\delta = \sum_{i=1}\big(\vert{\rm Re}(\nu_i)
- {\rm Re}(\nu_i^\prime)\vert + \vert{\rm Im}(\nu_i) - {\rm Im}(\nu_i^\prime)\vert\big)$, where real and
imaginary parts of $\nu_i$, $\nu_i^\prime$ are ordered before comparison. Figure~\ref{diagram}(a) shows
the distance $\delta$ in dependence on the coherence $\omega$ for the graphs in Fig.~\ref{graph}.
The $\omega$-spectra differ significantly over a wide range of the quantum-to-classical transition and no
specific level of coherence is required for distinguishing graphs. Interestingly, for the pairs in
Fig.~\ref{graph}(a) and (b), the DIP is maximal approximately halfway through the transition, indicating
that decoherence is essential for the performance of the algorithm.

The DIP of the $\omega$-spectrum compared to traditional spectra is summarized in Fig.~\ref{diagram}(b). Each
class in the diagram consists of all non-isomorphic pairs of graphs that are distinguishable by the spectrum
of either $A$, $L$ or $\gen^\omega$, as indicated. The important result is that the class of pairs distinguished
by the $\omega$-spectrum includes the entire classes defined by $A$ and $L$, and additional pairs not in these
classes. However, not all pairs are distinguishable by their $\omega$-spectra, as exemplified by strongly
regular graphs.


\section{Constructing $\boldsymbol\omega$-spectra from local fluctuations}

It is possible, in principle, to use the $\omega$-spectrum as the basis for quantum-inspired algorithms.
Such polynomial-time algorithms would involve the classical computation of the $\omega$-spectra of two graphs
and their subsequent comparison. However, we want to exploit the properties of QSWs in order to design a
quantum algorithm that relies on direct observations of QSWs and performs optimally in the presence of
decoherence. This is achieved by means of a counting device that monitors stochastic jumps of the QSW across
an auxiliary edge. Information about the graph structure is then encoded in the counting statistics $p(N)$ of
the number of jumps $N$ per time interval. As an essential part of our algorithm, we now explain how to
determine the $\omega$-spectrum from the counting statistics $p(N)$.

The counting statistics is obtained from the measuring device, consisting of the
auxiliary directed edge $(u\rightarrow v)$ with weight $\epsilon$, shown in Fig.~\ref{scheme}.
The vertices $u$ and $v$ are previously unconnected, that is $(u,v)\notin E$, but otherwise
chosen arbitrarily. Similar to regular edges, the auxiliary edge is described by the dissipator 
\begin{equation}\label{device}
\gen^{\rm aux}\rho = \epsilon\Big(\Upsilon_{uv}\rho\Upsilon_{uv}^\dagger
	- \frac{1}{2}\{\Upsilon_{uv}^\dagger \Upsilon_{uv},\rho\}\Big).
\end{equation}
The weight $\epsilon\ll 1$ is sufficiently small such that $\gen^{\rm aux}$ results in a negligible
perturbation of $\gen^\omega$ and $\sigma^\omega$.
The device measures fluctuations of random trajectories of the QSW in the steady state $\rho^{\rm ss}$,
specified by $(\gen^\omega + \gen^{\rm aux})\rho^{\rm ss}=0$ for $\omega\in[0,1]$. Fluctuations are
manifest in the number of stochastic jumps $N$ across the edge $(u\rightarrow v)$ during a fixed time
interval $\Delta t$. The random variable $N$, monitored by the device, is described by the counting
statistics $p(N)$ or equivalently the cumulants $C_{k} = \partial^k g(\chi)/\partial\chi^k\vert_{\chi=0}$,
with $g(\chi) = \log\av{\rme^{\chi N}}$ the cumulant generating function.

The $N$-resolved density matrix $\rho_N$ represents the QSW together with the counting device in state $N$.
The Laplace transform $\rho_\chi = \sum_N\rho_N\rme^{\chi N}$ is governed by the non trace preserving master
equation~\cite{garrahan2010thermodynamics,schaller2014open}
\begin{equation}\label{lindbladchi}
	\frac{\rmd \rho_\chi(t)}{\rmd t} =\gen^\omega(\chi)\rho_\chi(t)\,.
\end{equation}
The generator $\gen^\omega(\chi) = \gen^\omega + \gen^{\rm aux}(\chi)$ depends on $\chi$ to account
for the detector, where $\gen^{\rm aux}(\chi)$ is obtained from $\gen^{\rm aux}$ through the substitution
$\Upsilon_{uv}\rho\Upsilon_{uv}^\dagger\rightarrow \rme^\chi\Upsilon_{uv}\rho\Upsilon_{uv}^\dagger$.
The cumulant generating function $g(\chi)$ is then found from the solution $\rho_\chi(t)$ of Eq.~\eqref{lindbladchi}
in the long-time limit. Taking the trace over all states $\ket{i}$ yields
$\av{\rme^{\chi N}}=\tr[\rho_\chi(t)]\sim \rme^{\nu(\chi)t}$, with $\nu(\chi)$ the dominant eigenvalue
of $\gen^\omega(\chi)$, implying that $g(\chi)\sim\nu(\chi)t$ holds. Thus, for QSWs in the
steady state, the reduced cumulants $c_k\equiv C_k/\Delta t$ are given by
$c_k = \partial^k\nu(\chi)/\partial\chi^k\vert_{\chi=0}$.

Importantly, the cumulants $c_{k}$ are directly related to the coefficients $q_{i}(\chi)$ of the characteristic
polynomial $P^\omega(\nu,\chi) = \det[\nu\mathbb{1} - \gen^\omega(\chi)]$. The equality $P^{\omega}[\nu(\chi),\chi] = 0$
holds by definition for any $\chi$, such that by taking derivatives with respect to $\chi$ we obtain
the infinite set of equations~\cite{Bruderer-NJP-2014,wachtel2015fluctuating}
\begin{equation}\label{geneq}
	\frac{\rmd^\ell P^{\omega}[\nu(\chi),\chi]}{\rmd \chi^\ell}\bigg|_{\chi=0} = 0\qquad \ell = 1,2,\ldots
\end{equation}
involving $c_{k}$, $q_{i}(\chi)$ and their derivatives. We reveal structural details of Eqs.~\eqref{geneq}
by taking derivatives of individual monomials $q_i(\chi)\nu^i(\chi)$ of the polynomial $P^{\omega}[\nu(\chi),\chi]$,
yielding
\begin{equation}\label{polymer}
\begin{split}
	& \frac{\partial^\ell}{\partial\chi^\ell}q_i(\chi)\nu^i(\chi)\big\vert_{\chi=0} \\
	& = \sum_{k=0}^\ell\sum_{\vert p\vert=k}\binom{\ell}{k}\binom{k}{p}
\frac{\partial^{\ell-k}}{\partial\chi^{\ell-k}}q_i(\chi)\big\vert_{\chi=0}\,c_{p_1}\cdots c_{p_i},
\end{split}
\end{equation}
where $p = (p_1,\ldots,p_i)$ is an $i$-tuple of positive integers and $\vert p\vert = p_1 + \cdots + p_i$.
The most important observation is that Eqs.~(\ref{geneq}) are linear in the coefficients $q_i(\chi)$. Further,
all derivatives of $q_{i}(\chi)$ are identical to $q_{i}^\prime(\chi)$ due to the factor $\rme^\chi$ and
consequently Eqs.~(\ref{geneq}) depend on $q_{i}$ and $q_{i}^\prime\equiv q_{i}^\prime(\chi)\vert_{\chi=0}$ only.
Finally, because of the restriction $\vert p\vert=k$, we find that the $\ell$-th equation depends on cumulants
$c_{k}$ with order $k\leq\ell$ and coefficients $q_i$, $q^\prime_{i}$ with indices $i\leq \ell$ and $i\leq \ell-1$,
respectively.

Equations~(\ref{geneq}) can be utilized to construct the $\omega$-spectrum if sufficiently many
cumulants $c_k$ of the counting statistics $p(N)$ are known (see Ref.~\cite{Bruderer-NJP-2014} for details).
First note that Eqs.~(\ref{geneq}) involve $m=2(n^2-1)$ independent coefficients $q_1,\ldots,q_{n^2-1}$
and  $q_0^\prime,\ldots,q_{n^2-2}^\prime$ since $q_0 = q_{n^2-1}^\prime = q_{n^2}^\prime = 0$
and $q_{n^2}\equiv 1$. To determine the unknowns $q_i$, $q_i^\prime$ we set up a linear system
consisting of the first $\ell\leq m$ equations from~(\ref{geneq}), where cumulants $c_k$ with order $k\leq m$
enter as numerical factors. The linear system has a unique solution for the coefficients $q_i$, $q_i^\prime$,
from which the characteristic polynomial $P^{\omega}(\nu)$ and the desired $\omega$-spectrum are found.
Note that the coefficients $q_i$ and $\omega$-spectra are independent of the edge $(u\rightarrow v)$. By
contrast, the counting statistics $p(N)$, the cumulants $c_k$ and the coefficients $q_i^\prime$ depend on the
arbitrary choice of $(u\rightarrow v)$ and are therefore not suitable as graph invariants.

The cumulants $c_k$ are in practice approximated by unbiased minimum-variance estimators $\hat{c}_k$ obtained from
$s$ repeated observations of the number of jumps $N$ occurring during the interval $\Delta t$~\cite{kendall1946advanced}.
According to the Cram\'{e}r-Rao bound~\cite{prasad2004fisher}, their precision is quantified by the variance
$\var{\hat{c}_k}\sim k!c_2^k/s$, that is $s\sim k!$ for a prescribed precision. The observation time (i.e.~runtime)
necessary for determining $\omega$-spectra from cumulants of order $k\sim n^2$ therefore scales factorially with $n$,
and there is no speed advantage gained by the algorithm compared to conventional GI testing~\cite{corneil1980theoretical,mckay2014practical}.
Measuring cumulants with high precision seems challenging; however, this limitation might be overcome by
introducing several auxiliary edges.


\section{Conclusions}

We have presented a quantum algorithm for graph isomorphism (GI) testing that is
resilient to decoherence and performs optimal halfway through the quantum-to-classical transition.
Specifically, local observations of partially coherent quantum stochastic walks (QSWs) on graphs make
it possible to distinguish a large class of non-isomorphic graphs by means of their $\omega$-spectra.
While decoherence indeed improves the performance of our algorithm when compared to traditional
graph spectra we still have to clarify how powerful $\omega$-spectra are in comparison to other algorithms
by systematic benchmarking with larger sets of graphs~\cite{emms2009coined}.

Aside from GI testing, the $\omega$-spectrum contains further valuable information: The largest non-zero
real part of the $\omega$-spectrum determines the mixing time required for QSWs to reach the steady
state~\cite{mulken2011continuous}. Moreover, the $\omega$-spectrum is closely related to the algebraic
connectivity of the graph,~i.e.,~the smallest non-zero eigenvalue of the Laplacian~\cite{fiedler1973algebraic}.
For graphs of moderate size it is even practical to reconstruct the full graph from the spectrum by
using standard optimizations methods~\cite{comellas2008spectral}.

The presented methods for obtaining $\omega$-spectra are applicable to classical random walks and,
more generally, to open quantum systems~\cite{Bruderer-NJP-2014}. With this flexibility, we hope that
insights into quantum walks provided by our algorithm are transferable other algorithmic problems as
well as to physical and biological settings in which coherence plays an essential role.


\acknowledgements
The authors acknowledge useful discussions with O.~Marty, N.~Killoran and A.~Smirne.
This work has been supported by an Alexander von Humboldt Professorship,
the ERC Synergy grant BioQ and the EU projects SIQS and DIADEMS.


\bibliography{graph_walk_pra}

\end{document}